\documentclass{emulateapj}

\begin{document}

\shortauthors{Luhman \& Muench}
\shorttitle{Low-Mass Members of Chamaeleon~I}

\title{New Low-Mass Stars and Brown Dwarfs with Disks in the Chamaeleon~I 
Star-Forming Region\altaffilmark{1}}

\author{
K. L. Luhman\altaffilmark{2} \& A. A. Muench\altaffilmark{3}}

\altaffiltext{1}
{Based on observations performed with the Magellan Telescopes 
at Las Campanas Observatory and the {\it Spitzer Space Telescope}. 
}

\altaffiltext{2}{Department of Astronomy and Astrophysics, The Pennsylvania
State University, University Park, PA 16802; kluhman@astro.psu.edu.}

\altaffiltext{3}{Harvard-Smithsonian Center for Astrophysics, Cambridge, 
MA 02138.}

\begin{abstract}

We have used images obtained with the Infrared Array Camera and the 
Multiband Imaging Photometer onboard the {\it Spitzer Space Telescope}
to search for low-mass stars and brown dwarfs with circumstellar disks
in the Chamaeleon~I star-forming region.
Through optical spectroscopy of sources with red colors in these data,
we have identified seven new disk-bearing members of the cluster.
Three of these objects are probably brown dwarfs according to their spectral 
types (M8, M8.5, M8-L0).
Three of the other new members may have edge-on disks based on the shapes of
their infrared spectral energy distributions.
One of the possible edge-on systems has a steeply rising slope from 
4.5 to 24~\micron, indicating that it could be a class~I source 
(star+disk+envelope) rather than a class~II source (star+disk).
If so, then it would be one of the least massive known class~I protostars
(M5.75, $M\sim0.1$~$M_\odot$).

\end{abstract}


\section{Introduction}
\label{sec:intro}

Observations of circumstellar accretion disks around young stars provide
fundamental constraints on the processes of star and planet formation.
Such studies are facilitated by the identification of large, representative 
samples of disk-bearing members of star-forming regions.
The most obvious signature of a disk around a young star is the presence
of emission at infrared (IR) wavelengths in excess above that expected
from a stellar photosphere. 
One approach to applying this diagnostic has been to obtain mid-IR photometry
($\lambda\sim5$-20~\micron) for known members of star-forming regions 
\citep{kh95}. However, the resulting sample of stars is 
biased by the selection criteria that were originally used to discover
those objects. For instance, stars that are heavily obscured by edge-on disks,
protostellar envelopes, or molecular clouds are mostly
absent from optically-selected samples of young stars. 
Alternatively, wide-field mid-IR images of star-forming regions can be used 
to search for stars with disks in a relatively unbiased fashion.
The feasibility of such surveys has steadily improved over the past two
decades with advances in IR telescopes and detectors. 
The {\it Infrared Astronomical Satellite} ({\it IRAS}) imaged most of the sky
in four mid- and far-IR bands, producing the first comprehensive census of 
disks around young stars. However, because of its low spatial resolution,
{\it IRAS} was capable of resolving young stellar populations only 
in low-density regions like Taurus \citep{bei86,ken90}.
The better resolution and sensitivity of the
{\it Infrared Space Observatory} ({\it ISO}) allowed it to extend
mid-IR surveys to denser clusters and lower stellar
masses, including a few brown dwarfs \citep{per00,com98,com00,nt01,pas03}.
Ground-based cameras equipped with large-format IR detector arrays have 
offered even higher spatial resolution \citep{lada00,hai01}, but are 
restricted to the shortest IR bands where excess emission from disks is 
smaller ($\lambda\sim1$-4~\micron).

The {\it Spitzer Space Telescope} \citep{wer04} offers the best available
combination of field of view, sensitivity, spatial resolution, and
wavelength coverage for identifying young stars with disks 
\citep{all04,gut04,meg04,muz04}. 
The unique capabilities of {\it Spitzer} have been applied to a
large number of star-forming regions \citep[][references therein]{luh08cha}.
One of the primary objectives of these surveys is the extension of 
previous samples of disk-bearing stars to lower stellar masses. 
In an attempt to reach the lowest possible masses, {\it Spitzer} data have 
been used to search for new low-mass stars and brown dwarfs with disks 
in the nearest molecular clouds, including 
Taurus \citep{luh06tau2}, Perseus \citep{mue07}, Lupus \citep{all07},
Chamaeleon, and Ophiuchus \citep{allers06,allers07,luh05cha,luh08cha}.
We have continued this work by performing optical spectroscopy on
candidate low-mass objects in Chamaeleon~I 
\citep[$d=160$-170~pc,][]{whi97,wic98,ber99}.
In this paper, we describe the selection of these candidates from
{\it Spitzer} images (\S~\ref{sec:select}) and measure their optical 
spectral types 
(\S~\ref{sec:spectra}). We then characterize the stellar parameters, 
spatial distribution, and spectral energy distributions of the confirmed
members (\S~\ref{sec:prop}) and 
summarize their notable properties (\S~\ref{sec:disc}).
In the Appendix, we present measurements of {\it Spitzer} 
photometry for all known members of Chamaeleon~I that appear in 
the latest images of the region.

\section{Selection of Candidate Members of Chamaeleon~I}
\label{sec:select}

\begin{figure}
\epsscale{1}
\plotone{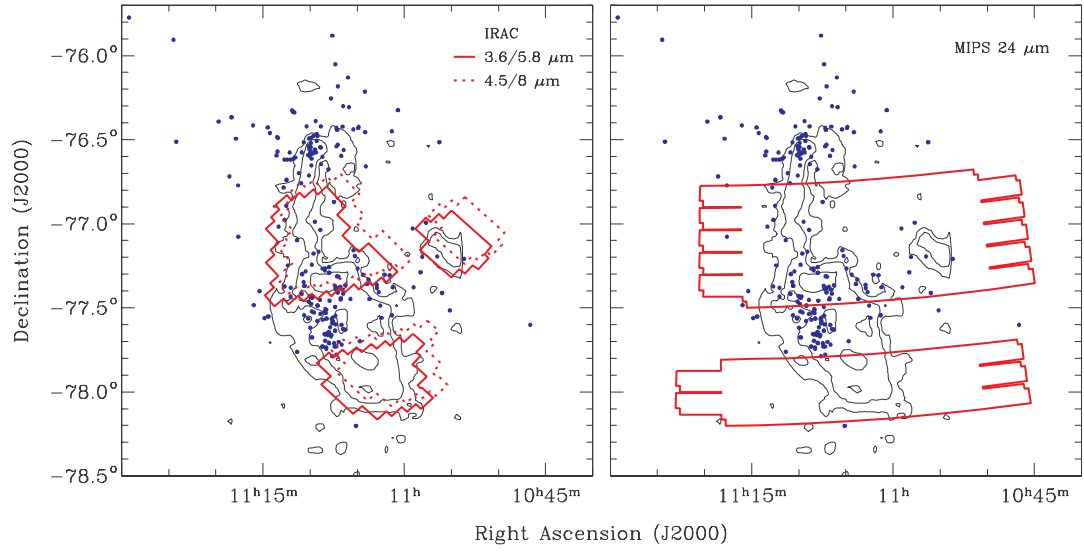}
\caption{
Fields in the Chamaeleon~I star-forming region that have been imaged
with IRAC ({\it left}) and MIPS ({\it right}) in {\it Spitzer} program 30574.
The known members of the cluster are indicated ({\it points}).
The contours represent the extinction map of \citet{cam97} at intervals
of $A_J=0.5$, 1, and 2.
}
\label{fig:map}
\end{figure}

To search for new disk-bearing members of Chamaeleon~I, we 
use images at 3.6, 4.5, 5.8, and 8.0~\micron\ obtained with
{\it Spitzer}'s Infrared Array Camera \citep[IRAC;][]{faz04} and
images at 24~\micron\ obtained with the Multiband Imaging Photometer for
{\it Spitzer} \citep[MIPS;][]{rie04}.
We consider all observations of this kind that have been performed in 
Chamaeleon~I, most of which were reduced and analyzed by \citet{luh08cha}. 
The photometric catalog produced by \citet{luh08cha} is used for this study.
The remaining observations of Chamaeleon~I that were not examined by 
\citet{luh08cha} were obtained through the {\it Spitzer} Legacy program of
L. Allen, which has a program identification of 30574.
The Astronomical Observation Request (AOR) identifications are
19986432, 19992832, 20006400, 20012800, 20014592, and 20015104 for the
IRAC observations and 
19978496, 19979264, 20010240, and 20011008 for the MIPS observations.
The IRAC and MIPS images were collected on 2007 May 15-17 and 2007 April 5
and were processed with the Spitzer Science Center (SSC) S16.1.0 and 
S16.0.1 pipelines, respectively.
We combined the images produced by the SSC pipeline into mosaics 
using R. Gutermuth's WCSmosaic IDL package. We then identified all point 
sources appearing in the resulting mosaics using the IRAF task STARFIND and
measured aperture photometry for them using the IRAF task PHOT. The details 
of these procedures are the same as those described by \citet{luh08cha}.
The total exposure times for the IRAC and MIPS images at a given 
position were 41.6 and $\sim$30~s, respectively. 
The boundaries of the IRAC and MIPS mosaics are indicated in the maps
of Chamaeleon~I in Figure~\ref{fig:map}.
In a given filter, the three IRAC mosaics cover areas of 0.10, 0.20, and 
0.31~deg$^2$. The two MIPS mosaics at 24~\micron\ encompass 1.4 and 
0.71~deg$^2$. The IRAC and MIPS photometric measurements from these 
images for known members of Chamaeleon~I are presented in the Appendix.

\citet{luh08cha} identified eight promising candidate members of Chamaeleon~I
in the {\it Spitzer} images that they analyzed. Those sources exhibit
red IRAC and MIPS colors that are indicative of circumstellar disks and
are located in the vicinity of known members of the star-forming region.
The candidacy of one of these objects, 2MASS~J11025374$-$7722561,
is also supported by its optical and near-IR colors \citep{lm04,luh07cha}.
To assess their membership, we selected for spectroscopy the six candidates 
that are bright enough for optical spectroscopy, consisting of 
2MASS J11020610$-$7718079, 2MASS J11025374$-$7722561, 
Cha J11062854$-$7618039, 2MASS J11085367$-$7521359, \\
2MASS J11100336$-$7633111 (also known as OTS~32), and 2MASS J11291470$-$7546256.
We also performed spectroscopy on 2MASS~J11095493$-$7635101, which was not
discussed by \citet{luh08cha}. It is a promising candidate because of its
red {\it Spitzer} colors and its close proximity to young stars in the
Cederblad~112 reflection nebula. 
Using the new {\it Spitzer} data from program 30574 that we have reduced
in this work, we searched for possible young stars with disks with the same 
criteria that were employed by \citet{luh08cha}, namely 
$[3.6]-[4.5]>0.15$, $[5.8]-[8.0]>0.3$, and errors less than 0.1~mag in 
all four bands.  From the resulting candidates, we selected 
2MASS J10533978$-$7712338 and Cha J11122701$-$7715173 for spectroscopy.
Finally, we included in our spectroscopic sample 2MASS J11091297$-$7729115,
which is the remaining bright candidate member appearing 
in the optical and near-IR color-magnitude diagrams from \citet{luh07cha}.

\section{Spectroscopy of Candidates}
\label{sec:spectra}

\subsection{Observations}

We obtained long-slit optical spectra of the 10 candidate members of 
Chamaeleon~I that were selected in \S~\ref{sec:select} using the 
Low Dispersion Survey Spectrograph (LDSS-3) on the Magellan~II Telescope
on the nights of 2007 December 17 and 18.
The spectra were taken through a 1.1$\arcsec$ slit. 
The instrument was operated with the VPH All and VPH Red grisms on
the first and second nights, respectively, resulting in 
spectral resolutions of 10 and 5.5~\AA\ at 7500~\AA.
All data were obtained with the slit rotated to the parallactic angle.
After bias subtraction and flat-fielding, the spectra were extracted and 
calibrated in wavelength with arc lamp data. The spectra were then corrected 
for the sensitivity functions of the detectors, which were measured from 
observations of a spectrophotometric standard star.

\subsection{Spectral Classification}

We now examine the LDSS-3 spectra for the evidence of youth that is 
expected for members of Chamaeleon~I.
The spectrum of 2MASS~J11291470$-$7546256 exhibits emission lines that
are indicative of a galaxy. Although the signal-to-noise ratio of its
data is low, we conclude that Cha~J11122701$-$7715173 is probably a 
background source rather than a low-mass member of the cluster based on 
the relatively blue slope and absence of late-type spectral features 
in its spectrum.
The remaining eight candidates do show spectral signatures
of young objects, such as strong H$\alpha$ emission and weak
Na~I and K~I absorption lines. We also detect He~I emission at 6678~\AA\ from
2MASS~J11085367$-$7521359 and Ca~II emission from OTS~32 and
Cha~J11062854$-$7618039.
Therefore, we classify these eight objects as members of Chamaeleon~I. 
The evidence of youth and membership is compiled in Table~\ref{tab:new}. 
When these new members are combined with the membership lists from 
\citet{luh07cha} and \citet{luh08cha}, the resulting census of Chamaeleon~I
contains 237 sources.

\begin{figure}
\epsscale{1}
\plotone{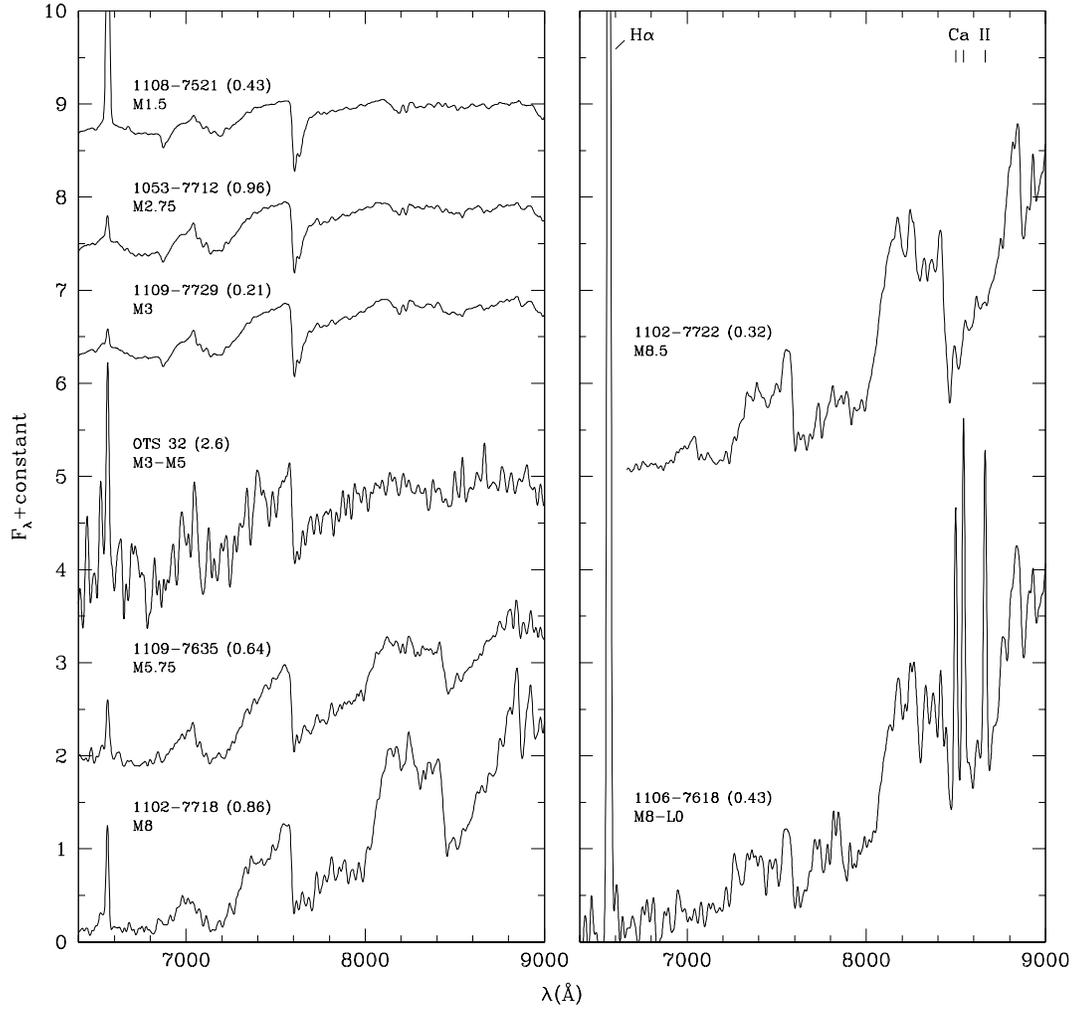}
\caption{
Optical spectra of new members of Chamaeleon~I.
The spectra have been corrected for extinction, which is quantified in
parentheses by the magnitude difference of the reddening
between 0.6 and 0.9~\micron\ ($E(0.6-0.9)$).
The data are displayed at a resolution of 18~\AA\ and are normalized at
7500~\AA.
}
\label{fig:spec}
\end{figure}

To measure spectral types for the eight new members, we have compared
their spectra to data for late-type members of 
Chamaeleon~I and other star-forming regions \citep{luh04cha,luh07cha},
which were originally classified at optical wavelengths through comparison 
to averages of dwarfs and giants \citep{luh99}.
The resulting classifications are presented in Table~\ref{tab:new}.
The spectra are shown in order of spectral type in Figure~\ref{fig:spec}.
The spectral type for Cha~J11062854$-$7618039 is more uncertain than
those of the other objects. A reasonable match to its spectrum is produced
by both M8 with $A_V\sim3$ and L0 with $A_V\sim1$. 
The latter agrees somewhat better
with the data, but a spectrum with a higher signal-to-noise ratio in
the TiO band near 7200~\AA\ is needed for a definitive classification.
A spectral type of L0 would make Cha~J11062854$-$7618039 tied with
Cha J11070768$-$7626326 \citep{luh08cha} as the coolest known member of 
Chamaeleon~I.

\section{Properties of New Members}
\label{sec:prop}

\subsection{Stellar Parameters}
\label{sec:par}

To examine the properties of the eight new members of Chamaeleon~I,
we begin by estimating their extinctions, effective temperatures, 
bolometric luminosities, and masses. 
Extinctions were derived from the optical spectra during the process
of spectral classification \citep{luh04cha,luh07cha}.
We have converted our spectral types to effective temperatures with the 
temperature scale from \citet{luh03ic}.
Luminosities have been estimated from $J$-band photometry in the manner
described by \citet{luh07cha}. Because near-IR data are unavailable 
for Cha~J11062854$-$7618039, we have estimated its $J$ magnitude by 
combining its 3.6~\micron\ measurement with the average value of
$J-[3.6]$ for late-type members of Chamaeleon~I that do not have mid-IR excess
emission \citep{luh08cha}. By doing so, we are assuming that the disk emission
at 3.6~\micron\ is negligible compared to the stellar photosphere, which
is true for most brown dwarfs with disks \citep{luh05cha}.
The uncertainties in $A_J$, $J$, BC$_J$, and the distance modulus 
($\sigma\sim0.13$, 0.05, 0.1, 0.13) correspond to total uncertainties
of $\pm0.09$ in log~$L_{\rm bol}$. 
The extinctions, temperatures, and luminosities for the
new members of Chamaeleon~I are listed in Table~\ref{tab:new}. 
We also include the available near-IR photometry for these objects.

\begin{figure}
\epsscale{1}
\plotone{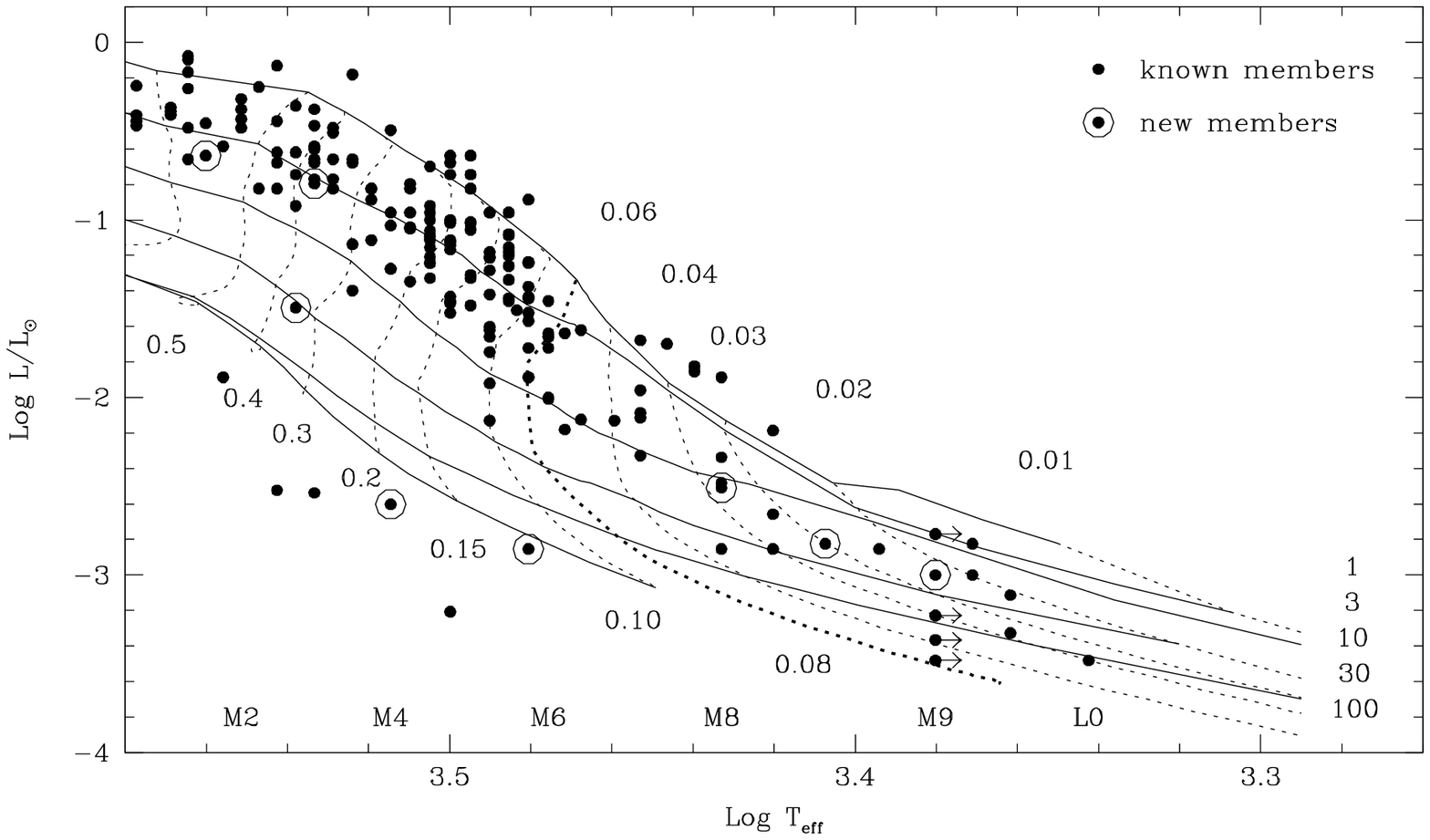}
\caption{
H-R diagram for all known low-mass stars and brown dwarfs in Chamaeleon~I
\citep[{\it points}]{luh07cha,luh08cha}. The sources classified as new members
through spectroscopy in Figure~\ref{fig:spec} are indicated ({\it circled
points}). These data are shown with the theoretical evolutionary models of
\citet{bar98} ($0.1<M/M_\odot\leq1$) and \citet{cha00} ($M/M_\odot\leq0.1$),
where the mass tracks ({\it dotted lines}) and isochrones ({\it solid lines})
are labeled in units of $M_\odot$ and Myr, respectively.
}
\label{fig:hr}
\end{figure}

The temperatures and luminosities of the new members are plotted on a 
Hertzsprung-Russell (H-R) diagram in Figure~\ref{fig:hr}. For comparison,
we also show the previously known low-mass members of Chamaeleon~I
\citep{luh07cha,luh08cha} and the predictions of theoretical evolutionary 
models \citep{bar98,cha00}. 
The positions of the three coolest new members in Figure~\ref{fig:hr} 
are within the sequence of known members and are indicative of substellar
masses. As discussed by \citet{luh08cha}, the precise value of the
mass estimate for a young late-type object depends on whether
it is derived from the temperature, the luminosity, or both.
The five other new members have masses ranging from 0.1 to 0.55~$M_\odot$ 
according to the data and models in Figure~\ref{fig:hr}. However, three of 
these objects, OTS~32, 2MASS J11095493$-$7635101, and 
2MASS J10533978$-$7712338, appear below the cluster sequence. 
The resulting isochronal ages ($\tau>30$~Myr) are unrealistically old 
considering that these sources exhibit clear signatures of youth 
($\tau<10$~Myr). These anomalously low 
luminosities may indicate that the stars are seen primarily in scattered light
(e.g., edge-on disks) at the shorter wavelengths from which the luminosities
were estimated. Indeed, we find that extended emission surrounds
2MASS~J11095493$-$7635101 in the near-IR images of Chamaeleon~I 
obtained by \citet{luh07cha}, supporting this hypothesis.
The nature of these sources is investigated further using their
spectral energy distributions in \S~\ref{sec:sed}.
We note that OTS~32 was originally identified as a possible member of 
Chamaeleon~I through the detection of $K$-band excess emission 
\citep{ots99,per99}. The faint near-IR magnitudes of this object were
suggestive of a substellar mass, but as we have shown, it is probably 
a low-mass star based on its mid-M spectral type.

\subsection{Spatial Distribution}

\begin{figure}
\epsscale{0.8}
\plotone{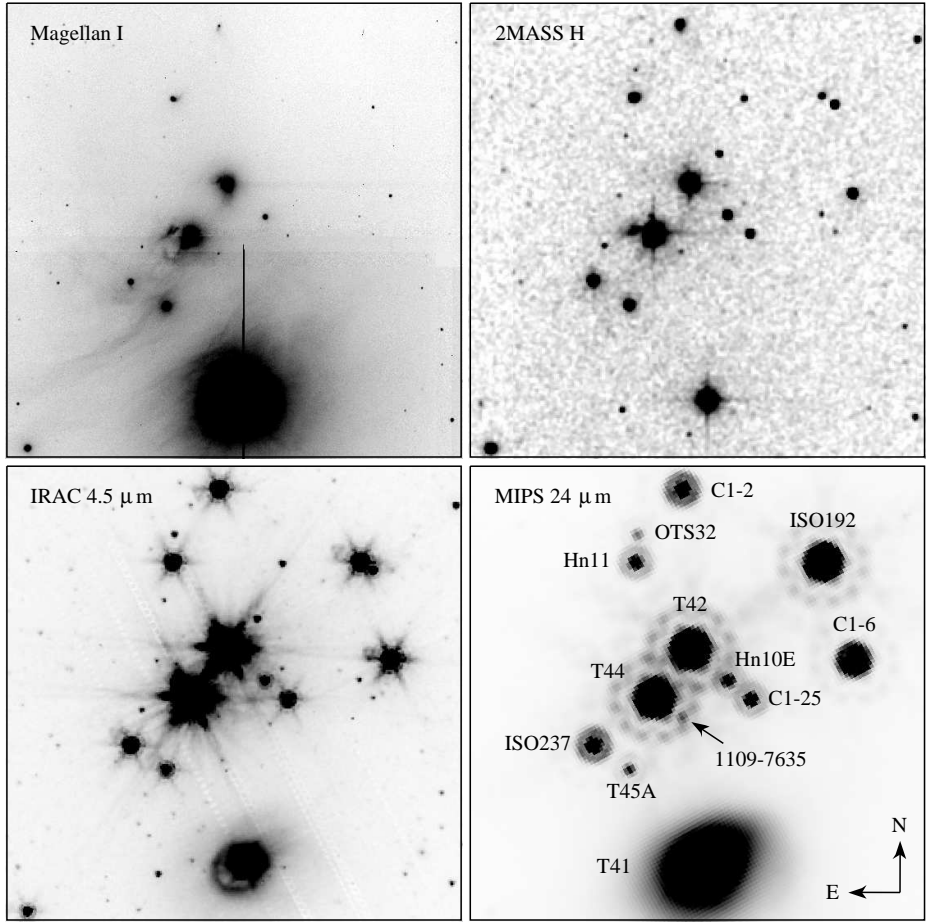}
\caption{
Optical and IR images of the region surrounding the Cederblad~112 
reflection nebula in Chamaeleon~I ($5\arcmin\times5\arcmin$).
All known members of the star-forming region within this area are labeled. 
OTS~32 and 2MASS J11095493$-$7635101 have been classified as new members 
through the spectra in Figure~\ref{fig:spec}.
}
\label{fig:image}
\end{figure}

The spatial positions of some of the new members of Chamaeleon~I merit 
discussion.
Two of these objects, OTS~32 and 2MASS~J11095493$-$7635101, are within
the group of young stars toward the Cederblad~112 reflection nebula,
as shown in the optical and IR images in Figure~\ref{fig:image}.
All of the 24~\micron\ sources within that $5\arcmin\times5\arcmin$ field
are spectroscopically confirmed members of the cluster.
Two of the new members, 2MASS~J11085367$-$7521359 and 2MASS~J10533978$-$7712338,
are relatively far from the bulk of the stellar population of Chamaeleon~I.
The former is $1\arcdeg$ north of the cloud complex
and is separated by $2\arcmin$ from the proper motion members 
RX~J1108.8$-$7519A and RX~J1108.8$-$7519B \citep{alc95,cov97,luh08cha}
and the latter is projected against a small cloudlet on the 
western edge of Chamaeleon~I.
Finally, 2MASS~J11091297$-$7729115 is only $4\arcsec$ from T39A and T39B. 
Thus, these stars may comprise a triple system.

\subsection{SED Classifications}
\label{sec:sed}

\begin{figure}
\epsscale{0.65}
\plotone{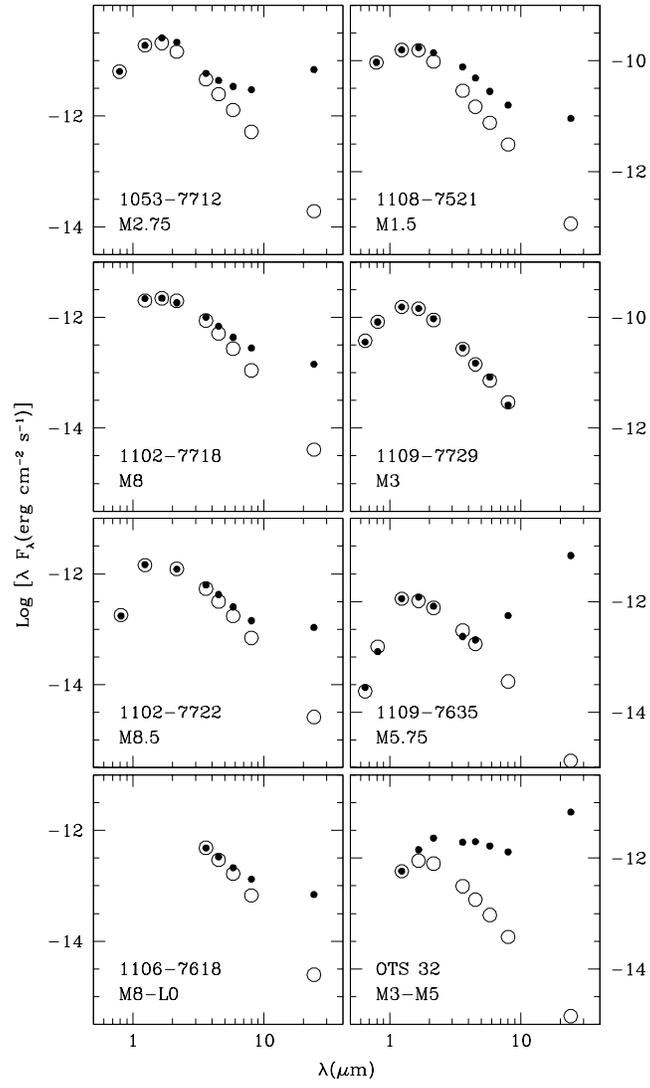}
\caption{
SEDs for new members of Chamaeleon~I ({\it points}).
Each object is compared to the SED of a stellar photosphere at the same 
spectral type ({\it circled points}). 
The photospheric SEDs have been reddened according to the extinction 
estimates in Table~\ref{tab:new} and have been scaled to the $J$-band 
fluxes of the new members, except for 2MASS J11020610$-$7718079 and 
Cha~J11062854$-$7618039,
for which the photospheres are scaled to $JHK_s$ and 3.6~\micron, respectively.
}
\label{fig:sed}
\end{figure}

All but one of the new members of Chamaeleon~I were selected for spectroscopy
in \S~\ref{sec:select} based on red mid-IR colors that suggested the 
presence of circumstellar disks. We now examine the IR spectral energy 
distributions (SEDs) of these sources in more detail.
To construct these SEDs, we use $I$-band photometry from \citet{lm04}, 
the Third Release of the Deep Near-Infrared Survey of the Southern Sky 
\citep[DENIS,][]{ep99}, and the Magellan IMACS images obtained by 
\citet{luh07cha}, near-IR photometry from 2MASS and 
\citet[][see Table~\ref{tab:new}]{luh07cha},
and mid-IR measurements from IRAC and MIPS that are listed in Table~6
from \citet{luh08cha} and in the Appendix. 
The resulting SEDs for the new members are plotted in Figure~\ref{fig:sed}.
To determine if long-wavelength excess emission is present, we compare each
SED to an estimate of the SED of the stellar photosphere, which is 
composed of the average colors of diskless stars near the spectral
type in question \citep{luh08cha}. The photospheric SEDs are reddened 
according to the extinctions in Table~\ref{tab:new} and the reddening
laws from \citet{rl85} and \citet{fla07} and are normalized to the 
$J$-band fluxes of the new members, except for 2MASS J11020610$-$7718079 
and Cha~J11062854$-$7618039. 
Because the available near-IR magnitudes of the former have large uncertainties,
the normalization is performed with the average of the $J$, $H$, and $K_s$
data. For Cha~J11062854$-$7618039, only IRAC and MIPS measurements are 
available. Therefore, we scale its photospheric template to the flux at 
3.6~\micron.  The emission from brown dwarfs with disks
at this wavelength is usually dominated by the photosphere, as 
discussed in \S~\ref{sec:par} and illustrated in the SEDs for the other 
late-type objects in Figure~\ref{fig:sed}.

As expected, the seven IR-selected members exhibit significant excess 
emission at long wavelengths relative to stellar photospheres.
The remaining member, 2MASS~J11091297$-$7729115, was identified as a 
candidate through optical and near-IR color-magnitude diagrams \citep{luh07cha}.
Its SED agrees well with that of a stellar photosphere and shows no evidence
of disk emission.
To characterize the SEDs quantitatively, we use spectral slopes defined as
$\alpha= d$~log$(\lambda F_\lambda)/d$~log$(\lambda)$ \citep{lw84,adams87}.
As in \citet{luh08cha}, we compute slopes between four pairs of bands, 
2.2-8, 2.2-24, 3.6-8, and 3.6-24~\micron. We deredden these slopes 
using the extinctions from Table~\ref{tab:new} and the reddening law 
from \citet{fla07}.
The resulting values of $\alpha_{2-8}$, $\alpha_{2-24}$, $\alpha_{3.6-8}$,
and $\alpha_{3.6-24}$ are presented in Table~\ref{tab:alpha}.
We also include the equivalent widths of the H$\alpha$ emission line measured
from our spectra.
We classify each object as class~I, flat-spectrum, class~II, or class~III 
by applying the thresholds from \citet{luh08cha} to the spectral
slopes, which follows the standard classification scheme for SEDs of 
young stars \citep{lada87,gre94}.

The SED classifications produced by $\alpha_{2-8}$, $\alpha_{2-24}$, 
$\alpha_{3.6-8}$, and $\alpha_{3.6-24}$ agree with each other
for five sources, but not for OTS~32, 2MASS~J11095493$-$7635101, 
and 2MASS~J10533978$-$7712338.
For the latter three objects, the SEDs become redder with longer wavelengths
such that the slopes ending at 8 and 24~\micron\ indicate class~II
and flat/class~I, respectively.
The distinctive behavior of these SEDs is indicative of stars that
are occulted by circumstellar material, such as edge-on disks
\citep{luh08cha}, which is consistent with the anomalously faint near-IR
magnitudes of these objects (\S~\ref{sec:par}). Indeed, the shape of 
the SED of OTS~32 closely resembles that of 2MASS~J04381486+2611399, 
which is a young brown dwarf in Taurus that has an edge-on disk 
\citep{luh07edgeon}. Thus, we tentatively classify OTS~32 and 
2MASS~J10533978$-$7712338 as class~II sources with edge-on disks.
Because 2MASS~J11095493$-$7635101 exhibits a more steeply rising SED
at 24~\micron, it could be a class~I source. 
The absence of excess emission at $\lambda<5$~\micron\ relative to
our estimate for its stellar photosphere may indicate the presence of an
inner cavity in its disk and envelope. Our SED classifications for
the new members are provided in Table~\ref{tab:alpha}.
Because the new class~I and class~II objects were identified as possible
members based on evidence of disks, they should
not be used in disk fraction measurements for Chamaeleon~I unless they 
are encompassed by the completeness limits of another survey for members 
that is unbiased in terms of disks.

\section{Discussion}
\label{sec:disc}

We conclude with a few remarks concerning notable aspects of
the eight new members of Chamaeleon~I that we have identified.
Three of the new members have spectral types later than M6, and thus
are likely to be brown dwarfs. The current census of Chamaeleon~I
now contains 33 known members later than M6. 
One of the new late-type members is classified as M8-L0; a type of L0 would 
make it one of the two coolest known members of the cluster and one of the 
least massive objects known to harbor a circumstellar disk. 
Additional optical spectroscopy 
of this object is needed for a more definitive spectral classification.
Seven of our new members were identified as possible members based on 
red mid-IR colors that indicated the presence of disks. 
As noted in \S~\ref{sec:intro}, a sensitive mid-IR survey 
of this kind is capable of finding disk-bearing young stars that are heavily 
obscured. For instance, because stars with edge-on disks are seen primarily 
in scattered light, they appear subluminous in optical color-magnitude 
diagrams and thus can be overlooked by optical surveys. 
Indeed, three of the new members exhibit properties that are indicative
of edge-on disks. High-resolution images and mid-IR spectroscopy are needed
to determine if edge-on disks are present \citep{luh07edgeon}.
One of the possible edge-on systems has a mass near the hydrogen burning mass 
limit according to its M5.75 spectral type and could be in
the class~I stage based on its rising SED from 4.5 to 24~\micron. 
In comparison, the coolest known class~I candidates prior to this work
were IRAS~04158+2805, IRAS~04248+2612, and IRAS~04489+3042 in Taurus 
\citep{kh95,whi04}, which have optical spectral types of M5.25, M4.5, and M4,
respectively \citep{luh06tau1}\footnote{These spectral types were derived in 
the same manner as the one for the new class~I source in Chamaeleon~I.
Slightly later spectral types for these three Taurus objects were reported by 
\citet{whi04}.}. Thus, this new object in Chamaeleon~I may be one of the least
massive known class~I sources.

\acknowledgements

K.~L. was supported by grant AST-0544588 from the National Science Foundation. 
This publication makes use of data products from 2MASS, which
is a joint project of the University of Massachusetts and the Infrared
Processing and Analysis Center/California Institute of Technology, funded
by NASA and the NSF.

\appendix

\section{{\it Spitzer} Photometry for Known Members of Chamaeleon I}
\label{sec:app}

\citet{luh08cha} presented {\it Spitzer} photometry for all members of
Chamaeleon~I that were known at that time and for a sample of candidate
members.
Those data were measured from all IRAC and MIPS 24~\micron\ images 
within $3\arcdeg$ of the star-forming region, with the exception of 
program 30574. As described in \S~\ref{sec:select}, we have reduced the 
data from program 30574 and used them to search for new members. 
In Table~\ref{tab:mem}, we present our IRAC and MIPS measurements for all 
known members appearing in those data. 
Five of the objects that we have confirmed as new members were presented
as candidates by \citet{luh08cha}. Thus, their {\it Spitzer} photometry is
provided in that study. One of the new members was identified in the
images from program 30574, and therefore is in Table~\ref{tab:mem}.
The remaining new members, 2MASS J11091297$-$7729115 and 
2MASS J11095493$-$7635101, were not in the lists of members and candidates in 
\citet{luh08cha}.
We include all available {\it Spitzer} measurements for these objects
in Table~\ref{tab:mem}. The IRAC data for 2MASS J11091297$-$7729115
were obtained on 2004 June 10 and the IRAC and MIPS data for
2MASS J11095493$-$7635101 were obtained on 2004 July 4 and 2004 April 11,
respectively. Table~\ref{tab:mem} and the tables in 
\citet{luh08cha} represent a compilation of all IRAC and MIPS 
24~\micron\ measurements for all known members of Chamaeleon~I.

We briefly summarize the implications of the new {\it Spitzer} photometry 
in Table~\ref{tab:mem} for our knowledge of the disk population in Chamaeleon~I.
Combining the data from \citet{luh08cha} and in Table~\ref{tab:mem},
IRAC and MIPS 24~\micron\ photometry has been measured for 208 and 166
members, respectively. Only eight of the 237 known members are outside
of all of the IRAC and MIPS images of this region.
Table~\ref{tab:mem} provides the first {\it Spitzer} photometry for 
eleven previously known members. As a result, the SEDs of these objects 
were not classified by \citet{luh08cha}. Based on our new photometry,
we classify T3A, T4, T5, T7, T8, T16, and T56 as class~II and 
2MASS~J11052272$-$7709290, Hn~7, Cam~2-42, and CHXR~57 as class~III.
\citet{luh08cha} did not compute spectral slopes like the ones in 
Table~\ref{tab:alpha} for T6, T27, and CHXR~54 because they were outside of the 
8 and 24~\micron\ images considered in that study.  As a result, they were 
classified using the IRAC data that were available at shorter wavelengths. 
Those classifications are confirmed by the full sets of IRAC and MIPS 
photometry for these stars that we now have in Table~\ref{tab:mem}.
T54 was outside of the IRAC images considered by \citet{luh08cha},
and was classified as class~II based on a 24~\micron\ measurement alone.
Our new IRAC data for this star from Table~\ref{tab:mem} 
do not exhibit excess emission and 
instead are consistent with emission from a stellar photosphere. 
This kind of SED in which excess emission suddenly appears at long
IR wavelengths is a signature of a disk with an inner hole, otherwise 
known as a transitional disk \citep{cal02,cal05,dal05,esp07a,esp07b,fur07}.
The presence of a transitional disk has been confirmed through
through spectroscopy with the {\it Spitzer} Infrared Spectrograph
(Furlan et al., in preparation). 
Finally, we find that T14A, ISO~91, T23, T35, and T47 have exhibited
significant variability (0.2-0.5~mag) between the IRAC and MIPS images from 
2004 that were analyzed by \citet{luh08cha} and the data from 2007 that are in 
Table~\ref{tab:mem}. In general, the magnitude change is similar among all
of the IRAC bands for a given object and pair of epochs.
All of these variable stars are class~I or class~II, which provides
additional evidence that stars with disks exhibit greater mid-IR variability 
than diskless stars \citep{liu96,bar05,luh08cha}.

\clearpage

\begin{deluxetable}{lllllllll}
\tabletypesize{\scriptsize}
\tablewidth{0pt}
\tablecaption{New Members of Chamaeleon I\label{tab:new}}
\tablehead{
\colhead{} &
\colhead{} &
\colhead{$T_{\rm eff}$\tablenotemark{b}} &
\colhead{} &
\colhead{$L_{\rm bol}$} &
\colhead{Membership} &
\colhead{} &
\colhead{} &
\colhead{} \\
\colhead{Name} &
\colhead{Spectral Type\tablenotemark{a}} &
\colhead{(K)} &
\colhead{$A_J$} &
\colhead{($L_\odot$)} &
\colhead{Evidence\tablenotemark{c}} &
\colhead{$J$\tablenotemark{d}} &
\colhead{$H$\tablenotemark{d}} &
\colhead{$K_s$\tablenotemark{d}}}
\startdata
2M J10533978$-$7712338 & M2.75 & 3451 & 0.63 & 0.032\tablenotemark{e} & ex,$A_V$ & 13.28$\pm$0.02 & 12.14$\pm$0.03 & 11.58$\pm$0.02 \\
2M J11020610$-$7718079 & M8 & 2710 & 0.56 & 0.0031 & e,ex,NaK & 15.63$\pm$0.08 & 14.81$\pm$0.07 & 14.26$\pm$0.09 \\  
2M J11025374$-$7722561\tablenotemark{f} & M8.5 & 2555 & 0.21 & 0.0015 & ex,NaK & 16.05$\pm$0.11 & \nodata & 14.70$\pm$0.13 \\
Cha J11062854$-$7618039 & M8-L0 & $\sim$2400 & $\sim$0.3 & $\sim$0.001 & e,ex,NaK & \nodata & \nodata & \nodata \\
2M J11085367$-$7521359\tablenotemark{g} & M1.5 & 3632 & 0.28 & 0.23 & e,ex & 10.98$\pm$0.03 & 10.07$\pm0.03$ & 9.56$\pm$0.03 \\ 
2M J11091297$-$7729115\tablenotemark{h} & M3 & 3415 & 0.14 & 0.16 & NaK & 11.00$\pm$0.04 & 10.27$\pm$0.03 & 9.98$\pm$0.03 \\
2M J11095493$-$7635101 &  M5.75 & 3024 & 0.42 & 0.0014\tablenotemark{e} & e,ex,NaK,$A_V$ & 16.34$\pm$0.03 & 15.47$\pm$0.02 & 15.13$\pm$0.03 \\
2M J11100336$-$7633111\tablenotemark{i} & M4$\pm$1 & 3270 & 1.7 & 0.0025\tablenotemark{e} & e,ex,$A_V$ & 17.06$\pm$0.02 & 15.29$\pm$0.02 & 14.01$\pm$0.02 \\ 
\enddata
\tablenotetext{a}{Uncertainties are $\pm0.25$ subclass unless noted otherwise.
These uncertainties represent the precision within the optical 
classification scheme adopted in this work \citep{luh99}.}
\tablenotetext{b}{Converted from the spectral types using the temperature 
scale from \citet{luh03ic}. In addition to the errors in the spectral type,
these temperature estimates are subject to a systematic uncertainty 
in the temperature scale \citep{luh08cha}, which is probably at least 
$\pm100$~K.}
\tablenotetext{c}{Membership in Chamaeleon is indicated by $A_V\gtrsim1$ and
a position above the main sequence for the distance of Chamaeleon (``$A_V$"),
strong emission lines (``e"), Na~I and K~I strengths intermediate
between those of dwarfs and giants (``NaK"), or IR excess emission (``ex").}
\tablenotetext{d}{From the ISPI images of \citet{luh07cha} for 
2M~J11095493$-$7635101 and 2M~J11100336$-$7633111 and from the
Point Source Catalog of the Two-Micron All-Sky Survey \citep[2MASS,][]{skr06}
for the remaining sources.}
\tablenotetext{e}{This star may have an edge-on disk. If so, this luminosity
estimate is not reliable.} 
\tablenotetext{f}{[LES2004] 424.}
\tablenotetext{g}{$2\arcmin$ from RX~J1108.8$-$7519A and RX~J1108.8$-$7519B.}
\tablenotetext{h}{$4\arcsec$ from T39A and T39B.}
\tablenotetext{i}{OTS 32. $18\arcsec$ from Hn 11.}
\end{deluxetable}

\begin{deluxetable}{lrrrrrr}
\tabletypesize{\scriptsize}
\tablewidth{0pt}
\tablecaption{H$\alpha$ and Spectral Slopes for New Members of Chamaeleon I\label{tab:alpha}}
\tablehead{
\colhead{} &
\colhead{$W_{\lambda}$(H$\alpha$)} &
\colhead{} &
\colhead{} &
\colhead{} &
\colhead{} &
\colhead{} \\
\colhead{Name} &
\colhead{(\AA)} &
\colhead{$\alpha$(2-8~\micron)} &
\colhead{$\alpha$(2-24~\micron)} &
\colhead{$\alpha$(3.6-8~\micron)} &
\colhead{$\alpha$(3.6-24~\micron)} &
\colhead{SED Class}
}
\startdata
2M J10533978$-$7712338 & 8$\pm$1 & $-$1.58 & $-$0.50 & $-$0.90 &  0.06 &  II \\
2M J11020610$-$7718079 & $\sim$100 & $-$1.49 & $-$1.09 & $-$1.64 & $-$1.04 &  II \\
2M J11025374$-$7722561 & $\sim$10 & $-$1.65 & $-$1.01 & $-$1.87 & $-$0.94 &  II \\
Cha J11062854$-$7618039 & $>$100 &    \nodata &    \nodata & $-$1.67 & $-$1.04 &  II \\
2M J11085367$-$7521359 & 82$\pm$3 & $-$1.68 & $-$1.14 & $-$2.01 & $-$1.14 &  II \\
2M J11091297$-$7729115 & 6.3$\pm$0.3 & $-$2.77 &    \nodata & $-$3.00 &    \nodata & III \\
2M J11095493$-$7635101 & 40$\pm$10 & $-$0.31 &  0.88 &  1.06 &  1.76 &   I or II \\
2M J11100336$-$7633111 & $\sim$20 & $-$0.64 &  0.36 & $-$0.61 &  0.62 &  II \\
\enddata
\end{deluxetable}

\begin{deluxetable}{lllllll}
\tabletypesize{\scriptsize}
\tablewidth{0pt}
\tablecaption{{\it Spitzer} Photometry for Known Members of 
Chamaeleon I\label{tab:mem}}
\tablehead{
\colhead{2MASS\tablenotemark{a}} &
\colhead{Name} &
\colhead{[3.6]} &
\colhead{[4.5]} &
\colhead{[5.8]} &
\colhead{[8.0]} &
\colhead{[24]} 
}
\startdata
J10533978$-$7712338 &         \nodata & 11.49$\pm$0.02 & 11.09$\pm$0.02 & 10.60$\pm$0.03 &  9.76$\pm$0.04 &  5.28$\pm$0.04 \\
J10555973$-$7724399 &        T3A &            out &            out &            out &            out &  2.37$\pm$0.04 \\
J10563044$-$7711393 &         T4 &  8.31$\pm$0.02 &            out &  7.58$\pm$0.03 &            out &  3.25$\pm$0.04 \\
J10574219$-$7659356 &         T5 &  8.74$\pm$0.02 &            out &  8.07$\pm$0.03 &            out &  4.35$\pm$0.04 \\
J10580597$-$7711501 &         \nodata &            out &            out &            out &            out &  7.21$\pm$0.05 \\
J10581677$-$7717170 &         T6 &            out &            out &            out &            out &  1.94$\pm$0.04 \\
J10590108$-$7722407 &         T7 &            out &            out &            out &            out &  3.45$\pm$0.04 \\
J10590699$-$7701404 &         T8 &            out &            out &            out &            out &  1.78$\pm$0.04 \\
J11013205$-$7718249 &    ESO H$\alpha$ 554 & 13.08$\pm$0.02 &            out & 12.92$\pm$0.04 &            out &        \nodata        \\
J11020610$-$7718079 &         \nodata & 13.43$\pm$0.02 &            out & 12.81$\pm$0.04 &            out &  9.49$\pm$0.22 \\
J11025374$-$7722561 & [LES2004] 424 &            out &            out &            out &            out &  9.65$\pm$0.18 \\
J11025504$-$7721508 &        T12 &            out &            out &            out &            out &  5.88$\pm$0.04 \\
J11034764$-$7719563 &        Hn 2 &  9.70$\pm$0.02 &            out &  9.53$\pm$0.03 &            out &  9.41$\pm$0.18 \\
J11035682$-$7721329 &     CHXR 12 &            out &            out &            out &            out &  9.31$\pm$0.15 \\
J11042275$-$7718080 &       T14A & 10.59$\pm$0.02 &  9.79$\pm$0.02 &  8.95$\pm$0.03 &  7.71$\pm$0.04 &  3.40$\pm$0.04 \\
J11044258$-$7741571 &      ISO 52 &            out &  9.81$\pm$0.02 &            out &  9.01$\pm$0.04 &      out          \\
J11045701$-$7715569 &        T16 &  9.80$\pm$0.02 &  9.57$\pm$0.02 &  9.18$\pm$0.04 &  8.63$\pm$0.04 &  6.11$\pm$0.04 \\
J11051467$-$7711290 &        Hn 4 &  9.26$\pm$0.02 &  9.16$\pm$0.02 &  9.12$\pm$0.03 &  9.18$\pm$0.04 &  8.92$\pm$0.11 \\
J11052272$-$7709290 &         \nodata & 11.48$\pm$0.02 & 11.39$\pm$0.02 & 11.38$\pm$0.03 & 11.37$\pm$0.04 &    \nodata            \\
J11054300$-$7726517 &     CHXR 15 &            out &            out &            out &            out &  9.09$\pm$0.14 \\
J11061540$-$7721567 &        Ced 110-IRS2/T21 &  6.25$\pm$0.02 &            out &  6.14$\pm$0.03 &            out &  5.58$\pm$0.07 \\
J11062942$-$7724586 &         \nodata &            out &            out &            out &            out &  8.70$\pm$0.20 \\
              \nodata &   Cha-MMS1 &             \nodata &             \nodata &             \nodata &             \nodata &  8.94$\pm$0.15 \\
J11063799$-$7743090 &    Cha H$\alpha$ 12 &            out & 11.21$\pm$0.02 &            out & 11.14$\pm$0.03 &  out              \\
J11064346$-$7726343 &        T22 &            out &            out &            out &            out &  8.86$\pm$0.17 \\
J11064510$-$7727023 &     CHXR 20 &            out &            out &            out &            out &  4.41$\pm$0.04 \\
J11064658$-$7722325 & Ced 110-IRS4 & 10.91$\pm$0.05 &            out &  9.23$\pm$0.04 &            out &  2.02$\pm$0.04 \\
J11065803$-$7722488 &      ISO 86 & 10.22$\pm$0.02 &            out &  8.27$\pm$0.03 &            out &  3.66$\pm$0.04 \\
J11065906$-$7718535 &        T23 &  9.76$\pm$0.02 &  9.41$\pm$0.02 &  8.96$\pm$0.03 &  7.99$\pm$0.04 &  5.07$\pm$0.04 \\
J11070369$-$7724307 &         \nodata &            out &            out &            out &            out &  7.30$\pm$0.05 \\
J11070919$-$7723049 & Ced 110-IRS6 &  8.00$\pm$0.02 &            out &  6.39$\pm$0.03 &            out &  1.64$\pm$0.04 \\
J11070925$-$7718471 &      ISO 91 & 10.56$\pm$0.02 & 10.15$\pm$0.02 &  9.76$\pm$0.03 &  8.98$\pm$0.04 &  5.35$\pm$0.04 \\
J11071148$-$7746394 &     CHXR 21 &            out &  9.25$\pm$0.02 &            out &  9.24$\pm$0.04 &            out    \\
J11071622$-$7723068 &      ISO 97 &  9.42$\pm$0.02 &            out &  8.24$\pm$0.03 &            out &  4.00$\pm$0.04 \\
J11072142$-$7722117 &        B35 &  9.55$\pm$0.02 &            out &  8.30$\pm$0.03 &            out &  3.55$\pm$0.04 \\
J11072825$-$7652118 &        T27 &  9.06$\pm$0.02 &  8.74$\pm$0.02 &  8.47$\pm$0.03 &  7.63$\pm$0.04 &  4.53$\pm$0.04 \\
J11073832$-$7747168 &    ESO H$\alpha$ 560 & 10.66$\pm$0.02 &            out & 10.56$\pm$0.03 &            out &     out           \\
J11075588$-$7727257 &     CHXR 28 &            out &            out &            out &            out &  7.29$\pm$0.04 \\
J11075730$-$7717262 &    CHXR 30B &  8.16$\pm$0.02 &  7.56$\pm$0.02 &  7.21$\pm$0.03 &  6.82$\pm$0.04 &  3.95$\pm$0.04 \\
J11075993$-$7715317 &    ESO H$\alpha$ 561 & 10.73$\pm$0.02 & 10.57$\pm$0.02 & 10.54$\pm$0.03 & 10.55$\pm$0.03 &      \nodata         \\
J11080002$-$7717304 &    CHXR 30A &  8.49$\pm$0.02 &  8.28$\pm$0.02 &  7.97$\pm$0.03 &  7.32$\pm$0.03 &  5.44$\pm$0.11 \\
J11082570$-$7716396 &         \nodata & 13.56$\pm$0.02 & 13.08$\pm$0.02 & 12.73$\pm$0.04 & 12.14$\pm$0.04 &  9.44$\pm$0.18 \\
J11082650$-$7715550 &     ISO 147 & 11.58$\pm$0.02 & 11.22$\pm$0.02 & 10.91$\pm$0.03 & 10.27$\pm$0.03 &  7.63$\pm$0.05 \\
J11083905$-$7716042 &        T35 &  8.71$\pm$0.02 &  8.45$\pm$0.02 &  8.27$\pm$0.03 &  8.21$\pm$0.04 &  4.75$\pm$0.04 \\
J11085464$-$7702129 &        T38 &  8.58$\pm$0.02 &  8.09$\pm$0.02 &  7.73$\pm$0.03 &  7.20$\pm$0.03 &  3.96$\pm$0.04 \\
J11090512$-$7709580 &        Hn 7 & 10.67$\pm$0.02 & 10.55$\pm$0.02 & 10.61$\pm$0.03 & 10.57$\pm$0.03 &          \nodata    \\
J11091297$-$7729115 & \nodata & 9.80$\pm$0.08 & 9.77$\pm$0.08 & 9.64$\pm$0.10 & 9.91$\pm$0.12 & \nodata \\
J11092913$-$7659180 & [LES2004] 602 & 11.75$\pm$0.02 & 11.62$\pm$0.02 & 11.56$\pm$0.03 & 11.53$\pm$0.04 &            \nodata    \\
J11093777$-$7710410 &    Cam 2-42 &  8.69$\pm$0.02 &  8.55$\pm$0.02 &  8.41$\pm$0.03 &  8.43$\pm$0.03 &  7.94$\pm$0.07 \\
J11094260$-$7725578 &       C7-1 &            out &            out &            out &            out &  6.55$\pm$0.04 \\
J11094742$-$7726290 &        B43 &            out &            out &            out &            out &  4.41$\pm$0.04 \\
J11094866$-$7714383 &     ISO 209 & 11.28$\pm$0.02 & 10.82$\pm$0.02 & 10.46$\pm$0.03 &  9.85$\pm$0.03 &  7.25$\pm$0.04 \\
J11095336$-$7728365 &     ISO 220 &            out &            out &            out &            out &  7.25$\pm$0.05 \\
J11095493$-$7635101 & \nodata & 15.00$\pm$0.13 & 14.43$\pm$0.23 &         \nodata & 11.59$\pm$0.08 &  5.30$\pm$0.20 \\
J11100785$-$7727480 &     ISO 235 &            out &            out &            out &            out &  6.36$\pm$0.04 \\
J11102852$-$7716596 &      Hn 12W & 10.37$\pm$0.02 & 10.25$\pm$0.02 & 10.26$\pm$0.03 & 10.22$\pm$0.04 &           \nodata     \\
J11103481$-$7722053 & [LES2004] 405 &  9.65$\pm$0.02 &  9.52$\pm$0.02 &  9.42$\pm$0.03 &  9.48$\pm$0.03 &  9.06$\pm$0.11 \\
J11103644$-$7722131 &     ISO 250 & 10.15$\pm$0.02 & 10.07$\pm$0.02 &  9.98$\pm$0.03 &  9.94$\pm$0.04 &  9.61$\pm$0.12 \\
J11104141$-$7720480 &     ISO 252 & 11.45$\pm$0.02 & 11.02$\pm$0.02 & 10.59$\pm$0.03 &  9.75$\pm$0.03 &  7.00$\pm$0.04 \\
J11104959$-$7717517 &        T47 &  8.92$\pm$0.02 &  8.35$\pm$0.02 &  7.76$\pm$0.03 &  6.63$\pm$0.04 &  3.00$\pm$0.04 \\
J11105076$-$7718031 &    ESO H$\alpha$ 568 & 10.38$\pm$0.02 & 10.28$\pm$0.02 & 10.22$\pm$0.03 & 10.17$\pm$0.03 &    \nodata            \\
J11105359$-$7725004 &     ISO 256 &  9.67$\pm$0.02 &            out &  8.70$\pm$0.03 &            out &  4.90$\pm$0.04 \\
J11105597$-$7645325 &       Hn 13 &            out &  9.10$\pm$0.02 &            out &  8.02$\pm$0.03 &           out     \\
J11112260$-$7705538 &     ISO 274 & 10.43$\pm$0.02 & 10.35$\pm$0.02 & 10.29$\pm$0.03 & 10.30$\pm$0.03 &             \nodata   \\
J11120288$-$7722483 &         \nodata & 12.10$\pm$0.02 & 12.00$\pm$0.02 & 11.92$\pm$0.03 & 11.94$\pm$0.04 &           \nodata    \\
J11120351$-$7726009 &     ISO 282 & 10.83$\pm$0.02 &            out & 10.13$\pm$0.03 &            out &  6.87$\pm$0.05 \\
J11122250$-$7714512 &         \nodata & 13.89$\pm$0.02 &            out & 13.55$\pm$0.05 &            out &         \nodata       \\
J11123099$-$7653342 & [LES2004] 601 & 12.59$\pm$0.02 & 12.47$\pm$0.02 & 12.39$\pm$0.04 & 12.36$\pm$0.05 &             \nodata  \\
J11124210$-$7658400 &     CHXR 54 &  9.41$\pm$0.02 &  9.43$\pm$0.02 &  9.35$\pm$0.03 &  9.36$\pm$0.03 &  9.35$\pm$0.11 \\
J11124268$-$7722230 &        T54 &  7.81$\pm$0.02 &  7.83$\pm$0.02 &  7.75$\pm$0.03 &  7.66$\pm$0.03 &  4.85$\pm$0.04 \\
J11124861$-$7647066 &       Hn 17 &            out &            out &            out &            out &  7.01$\pm$0.04 \\
J11132012$-$7701044 &     CHXR 57 &  9.79$\pm$0.02 &            out &  9.74$\pm$0.03 &            out &  9.33$\pm$0.15 \\
J11173700$-$7704381 &        T56 &            out &            out &            out &            out &  4.22$\pm$0.04 \\
\enddata
\tablecomments{Entries of ``$\cdots$" and ``out"
indicate measurements that are absent because of non-detection
and a position outside the field of view of the filter for the observations
in program 30574, respectively, except for 2MASS J11091297$-$7729115. 
For this star, the non-detection at 24~\micron\ refers to images from 
program 37 obtained on 2005 February 27.}
\tablenotetext{a}{2MASS Point Source Catalog.}
\end{deluxetable}

\clearpage

\end{document}